\documentclass[prb,reprint,twocolumn,superscriptaddress,noshowpacs,notitlepage,longbibliography,10pt,citeautoscript]{revtex4-2}%

\usepackage{graphicx,bm,times}
\usepackage{amsmath}
\usepackage{amsfonts}
\usepackage{amssymb}
\usepackage{mathtools}
\usepackage{color}
\usepackage[normalem]{ulem} 
\usepackage{hyperref}

\hypersetup{
	colorlinks = true,
	allcolors = {blue}
}

\begin{document}

\title{Clustering-Enhanced Time- and Angle-Resolved Photoemission Study of LaTe$_3$: Absence of a Photoinduced Secondary CDW in the Electronic Structure}

\author{Gesa-R. Siemann}
\affiliation{Department of Physics and Astronomy, Aarhus University, 8000 Aarhus C, Denmark}
\author{Davide Curcio}
\affiliation{CNR, Consiglio Nazionale delle Ricerche, Istituto officina dei materiali, Trieste, Italy}
\author{Anders S. Mortensen}
\affiliation{Department of Physics and Astronomy, Aarhus University, 8000 Aarhus C, Denmark}
\author{Charlotte E. Sanders}
\affiliation {Central Laser Facility, STFC Rutherford Appleton Laboratory, OX11 0QX, Harwell, UK}
\affiliation{SUPA, School of Physics and Astronomy, University of St Andrews, North Haugh, St Andrews, UK}
\author{Yu Zhang}
\affiliation {Central Laser Facility, STFC Rutherford Appleton Laboratory, OX11 0QX, Harwell, UK}
\author{Jennifer Rigden}
\affiliation {Central Laser Facility, STFC Rutherford Appleton Laboratory, OX11 0QX, Harwell, UK}
\author{Paulina Majchrzak}
\affiliation{Department of Applied Physics, Stanford University, Stanford, CA, USA}
\author{Deepnarayan Biswas}
\affiliation{Diamond Light Source Ltd, Harwell Science and Innovation Campus, Didcot, UK}
\author{Emma Springate}
\affiliation {Central Laser Facility, STFC Rutherford Appleton Laboratory, OX11 0QX, Harwell, UK}
\author{Ratnadwip Singha}
\affiliation{Department of Chemistry, Princeton University, Princeton, USA}
\affiliation{Department of Physics, Indian Institute of Technology Guwahati, Assam 781039, India}
\author{Leslie M. Schoop}
\affiliation{Department of Chemistry, Princeton University, Princeton, USA}
\author{Philip Hofmann}
\email{email: philip@phys.au.dk}
\affiliation{Department of Physics and Astronomy, Aarhus University, 8000 Aarhus C, Denmark}

\date{\today}

\begin{abstract} 
Optical control offers a compelling route for tailoring material properties on an ultrafast time scale. Ordered states such as charge density waves (CDWs) can be transiently melted by an ultrafast light excitation. This is also the case for the rare-earth tritelluride LaTe$_3$, a prototypical CDW compound. For this material it has recently been reported that the suppression of the primary CDW allows the transient formation of a second CDW, whose wave vector is orthogonal to the primary one. This creates the intriguing scenario where light enables switching between two distinct ordered phases of the material. While the second CDW has so far been observed by structural techniques, it remains an open question how the interplay of the two CDW phases is reflected in the material's electronic structure. We investigate this via time- and angle-resolved photoemission measurements of LaTe$_3$. The complex Fermi contour is probed using a FeSuMa analyzer, which records the photoemission intensity of the entire Fermi contour at once. The dynamics revealed by the FeSuMa analyzer are complemented by measurements using a conventional hemispherical electron analyzer. We combine conventional data analysis with $k$-means clustering, an unsupervised machine learning technique, demonstrating its strong potential for disentangling large datasets. While we do not find any features that cannot be explained by the melting and re-establishment of the primary CDW, distinct dynamics and coherent oscillations are observed in the different branches of the Fermi contour. 
\end{abstract}

\maketitle

\section*{Introduction}
 The rare-earth tritellurides $R$Te$_3$, where $R$ represents a rare-earth element, have attracted significant interest as prototype materials for a nesting-driven, quasi-one-dimensional charge density wave (CDW). The compounds crystallize in a quasi-tetragonal lattice with space group No. 63 (Cmcm). The small in-plane anisotropy between the $c$- and $a$-axes energetically favours the formation of an incommensurate CDW along the former \cite{dimasiChemicalPressureChargedensity1995,gweonDirectObservationComplete1998,malliakasSquareNetsTellurium2005, laverockFermiSurfaceNesting2005,brouetAngleresolvedPhotoemissionStudy2008}. The transition temperature of this CDW phase gradually increases for lighter $R$ elements~\cite{ruEffectChemicalPressure2008,malliakasDivergenceBehaviorCharge2006, huCoexistenceCompetitionMultiple2014}. For heavier $R$ elements, on the other hand, the smaller in-plane lattice constant results in a reduction of the effective in-plane anisotropy, giving rise to a second CDW along the $a$-axis~\cite{ruMagneticPropertiesCharge2008,mooreFermiSurfaceEvolution2010,banerjeeChargeTransferMultiple2013, huCoexistenceCompetitionMultiple2014}.
La is the lightest atom in the series of rare-earth elements, and LaTe$_3$ thus exhibits the largest in-plane lattice constant, the largest CDW gap, and an estimated transition temperature of around 670~K \cite{huCoexistenceCompetitionMultiple2014}. The lightest $R$ element known to give rise to an equilibrium CDW along the $a$-axis is Tb \cite{banerjeeChargeTransferMultiple2013} while this second CDW has never been observed in LaTe$_3$ under equilibrium conditions.

\begin{figure}[t]
\includegraphics[scale=0.75]{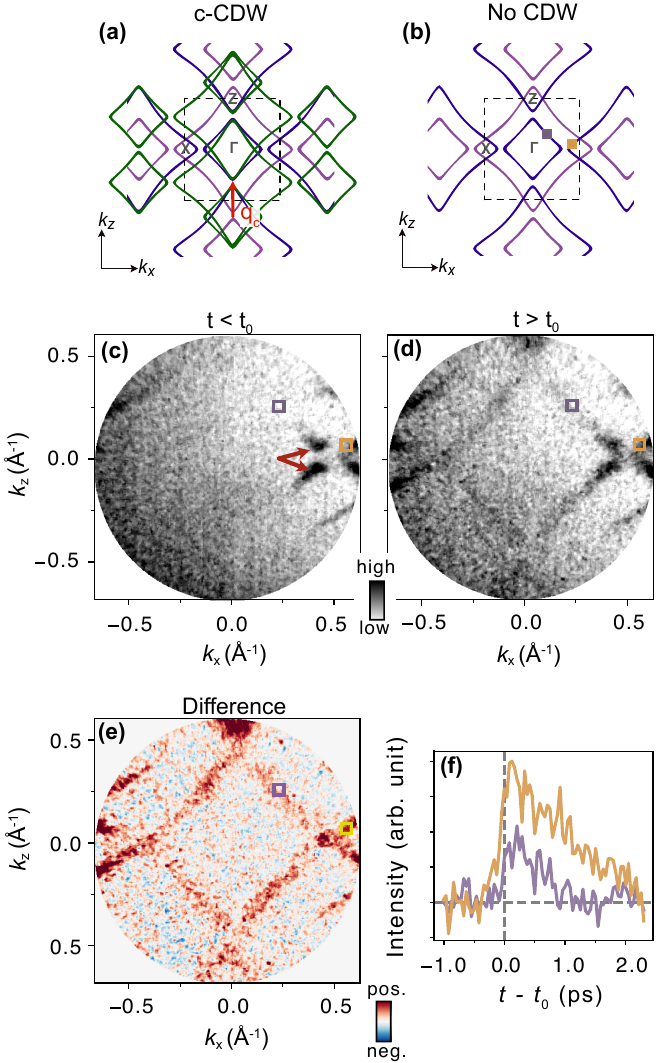}
\caption{(a) Illustration of CDW gap formation via Fermi-contour replicas generated by the periodic lattice distortion with wave vector $q_c$. A gap opens where the original Fermi contour overlaps with a replica.
(b) Sketch of the LaTe$_3$ Fermi contour in the absence of a CDW. The dashed square marks the surface-projected Brillouin zone.
(c) Equilibrium Fermi contour measured with the FeSuMa analyzer. The red arrows mark the so called butterfly structure.
(d) Fermi contour near peak excitation ($t - t_0 = 170$~fs).
(e) Difference between the data in (c) and (d). Red indicates a positive and blue a negative difference.
(f) Photoemission intensity as a function of time delay, integrated over the two regions marked in panels (b)–(e) by the purple and yellow squares.
}
\label{fig:fig1}
\end{figure}

Surprisingly, recent ultrafast electron diffraction experiments have shown that a transient CDW along the $a$-axis can be observed in LaTe$_3$ upon suppression of the $c$-axis CDW by an ultrafast laser pulse \cite{Kogar:2019ab, Zong:2021aa}.  
This observation of two distinct CDW phases in the structural measurements raises the question of how the interplay between the two CDWs and their respective gaps is reflected in the transient electronic structure. The most direct approach to address this is time- and angle-resolved photoemission spectroscopy (TR-ARPES) \cite{Sobota:2021wy,Boschini:2024aa}.
A number of TR-ARPES studies of LaTe$_3$ \cite{Zong:2018ab, liuDistinctAmplitudeMode2024}, as well as some of its sister compounds, have been reported \cite{Schmitt:2008aa, Schmitt:2011aa, rettigCoherentDynamicsCharge2014, maklarCoherentModulationQuasiparticle2022, Trigo:2021aa, Maklar:2021aa, suTimedomainIdentificationDistinct2025}. All of these studies reveal clear signatures of the $c$-CDW phase melting upon light excitation, with some also reporting coherent oscillations of the CDW amplitude mode. Crucially, none of these studies report any signature of a second $a$-CDW in the non-equilibrium electronic structure of LaTe$_3$.

However, the presence of a second $a$-CDW has only been considered very recently, and a possible reason for not having observed it in TR-ARPES may be that most of the studies focused on regions in $k$-space associated with the $c$-CDW, such as the area of largest gap opening along the $k_z$ direction. Moreover, studies of the sister compounds have shown that the electronic structure signatures of the $a$-CDW are rather subtle \cite{mooreFermiSurfaceEvolution2010}, at least in the presence of a simultaneous $c$-CDW. Consequently, the manifestations of an $a$-CDW in the transient electronic structure of LaTe$_3$ could be  weak, requiring careful analysis to disentangle the melting of the primary CDW phase from the emergence of a secondary one.

The objective of this study is to establish the electron dynamics of photoexcited LaTe$_3$, particularly near the Fermi energy, by performing TR-ARPES measurements across the entire Brillouin zone (BZ), rather than focusing on selected regions. This approach is intended to not only address whether the second $a$-CDW can be detected in the transient electronic structure, but also to reveal potential variations in electron dynamics across different parts of the Fermi surface.

The technical challenges of this endeavor are twofold. First, comparing the dynamics throughout the BZ requires a spectrometer capable of \emph{simultaneously} collecting data over a large portion of $k$-space—such as a momentum microscope \cite{Medjanik:2017aa, Kutnyakhov:2020aa} or, as used in the present case, a FeSuMa retarding field analyzer \cite{borisenkoFermiSurfaceTomography2022, majchrzakAccessFullThreedimensional2024}. Second, it is essential to systematically analyze the electron dynamics across the full Fermi surface in order to detect possible changes and distinct behavior of the electrons after excitation. To this end, we apply $k$-means clustering—an unsupervised machine learning technique that has proven effective in identifying subtle trends in large TR-ARPES datasets \cite{meyerLineShapesTime2025, majchrzakMachinelearningApproachUnderstanding2025a}.

Despite the $R$Te$_3$ materials being model systems for a nesting-driven, quasi-one-dimensional CDW, their electronic structure is rather complex. The quasi-two-dimensional Fermi contour (FC) is similar across all $R$Te$_3$ compounds, as it predominantly derives from the in-plane $p$ orbitals of the quasi-square-shaped Te sheets. A detailed account of the FC and its relation to the CDW can be found in Refs.~\cite{brouetFermiSurfaceReconstruction2004, brouetAngleresolvedPhotoemissionStudy2008, chikinaChargeDensityWave2023}; here, we provide only a brief summary.
A sketch of the FC in the non-CDW state is shown in Fig.~\ref{fig:fig1}(b). It results from interacting $p_x$ and $p_z$ orbitals, combined with back-folding due to the three-dimensional crystal structure. In the sketch, the main FC elements are shown in blue, while the back-folded ones appear in purple. In the following, we distinguish between the “inner square” part of the FC—the closed blue contour around $\Gamma$—and the “outer” FC, which derives from the corners of the larger squares that extend into neighboring Brillouin zones, forming the small pockets around the X and Z-points. The nearly square symmetry of the material is also reflected in the FC, which, in this simplified picture, exhibits fourfold symmetry. 

The formation of the $c$-CDW is illustrated in Fig.~\ref{fig:fig1}(a). The periodic lattice distortion associated with the CDW wave vector $q_{c}$ generates the green FC replicas. These replica bands overlap almost perfectly with the original FC in the first Brillouin zone along the $\Gamma-Z$ direction, leading to gap formation around the Fermi energy $E_\mathrm{F}$. In ARPES, this results in a suppression of photoemission intensity at $E_\mathrm{F}$ along the entire $\Gamma$–Z corridor of the BZ. What remains visible are the structures near the X point, derived from both the outer FC and the corners of the inner square. These regions are made out of a number of crossings of replica and original bands and as a consequence exhibit considerable  complexity. This remaining feature near X is often referred to as the butterfly structure (see also red arrows in Fig.~\ref{fig:fig1}(c)). 

The formation of a second CDW along the $a$-axis could be described similarly: replicas should appear, displaced by the corresponding CDW wave vector $q_a$, leading to additional gap openings at $E_\mathrm{F}$ now along the $\Gamma$-X direction. For a single $c$-CDW, the experimental results are well described by a tight-binding model \cite{brouetAngleresolvedPhotoemissionStudy2008, chikinaChargeDensityWave2023}, while a similar approach captures the coexistence of both $a$- and $c$-CDWs in equilibrium \cite{mooreFermiSurfaceEvolution2010}. In ARPES, CDW effects can be detected either through the gap openings or via the presence of replica bands \cite{brouetFermiSurfaceReconstruction2004, brouetAngleresolvedPhotoemissionStudy2008, sarkarChargeDensityWave2023}.

\section*{Methods}

The TR-ARPES data presented in this study were measured at the Artemis laboratory of the Central Laser Facility in the UK. The samples were top-posted and cleaved $in$-$situ$ to obtain a clean surface at a base pressure better than 9$\times 10^{-10}$~mbar. Both the pump and probe beams were generated from the same 100~kHz laser system~\cite{thireVersatileHighaveragepowerUltrafast2023}, with their respective linear polarizations being perpendicular and parallel to the plane of incidence. All experiments were performed at a sample temperature of approximately 80~K using liquid nitrogen cooling.  

The electron dynamics near the FC was measured using a FeSuMa analyzer~\cite{borisenkoFermiSurfaceTomography2022, majchrzakAccessFullThreedimensional2024}. This is a retarding field analyzer where all electrons with a kinetic energy lower than the retarding potential are repelled. The measured spectrum therefore corresponds to the integrated intensity of all electrons with kinetic energies higher than the applied retarding potential. In order to measure the FC, the retarding voltage is typically set to just below the kinetic energy corresponding to $E_\mathrm{F}$. Note that at finite temperature or in the case of strong pump excitation, electrons above $E_\mathrm{F}$ are also detected by the FeSuMa. However, this signal is usually quite small compared to the dominant intensity from the fully populated states below $E_\mathrm{F}$. In order to evaluate the possible contribution of such hot electrons, we present a data set with the retarding voltage just below $E_\mathrm{F}$ (within 25~meV) in the main text of this paper and provide a second data set with the retarding voltage approximately 100~meV below $E_\mathrm{F}$ in the Appendix. We find that the difference between the data sets is very small, excluding a major effect of hot electrons on the results.  

For the FeSuMa measurements the pump and probe photon energy was 1.42~eV and 28.4~eV, respectively, with a fluence of 5~mJ/cm$^2$. In order to increase the angular acceptance and to be able and measure a large portion of the Fermi surface an additional ``fisheye" voltage of 28~eV was applied during the data acquisition. The FeSuMa analyzer has an angluar resolution of 0.2$^{\circ}$, resulting in a momentum resolution on the order of $\Delta k \approx 0.01$~\AA$^{-1}$; however, the real momentum resolution depends also on the energy resolution, in particular for a retarding field analyzer. The optimal energy resolution of the analyzer is 12~meV, but this is much smaller than the actual energy resolution of a pump-probe experiment like this one, which is limited by the Fourier transform limit and space charge effects. For the case of this experiment, the Fourier transform limit of a 60-fs probe pulse would be $\Delta E \approx$ $45$~meV. 

The FeSuMa measurements were complemented by data collected with a conventional hemispherical analyzer along selected directions in $k$-space. During this data acquisition the probe photon energy was $h\nu = 39~$eV and the fluence 4~mJ/cm$^2$. For the hemispherical analyzer, the energy resolution was limited by the analyzer settings resulting in $\Delta E \approx 110$~meV and the momentum resolution was $\Delta k = 0.02$~\AA$^{-1}$. 

The time resolution is a convolution of the temporal widths of the pump and probe pulses.  The intrinsic pulse width of the driving laser for both pump and probe is 60~fs.  Overall, across the several measurements and different measurement conditions described in this paper, the pump pulse ranged from 60 to 75~fs in width.  The beamline does not have a direct measurement of the temporal width of the probe pulse, but the expected wavefront distortion induced by the gratings would lead to a width of approximately 90 to 125~fs for data acquired at 39~eV, and 120 to 170~fs for data acquired at 28.4~eV.

To identify regions of different dynamics, the experimental data were analyzed using $k$-means clustering, an unsupervised machine learning technique~\cite{ballClusteringTechniqueSummarizing1967, MacQueen:1967aa, xuClustering2008}. We largely follow the approach described in Ref.~\cite{majchrzakMachinelearningApproachUnderstanding2025a}. Specifically, the clustering algorithm was applied to the time-dependent photoemission intensity in a region of interest (ROI) in either the constant-energy surface measured by the FeSuMa or in the energy vs.~$k$ distribution mapped by the hemispherical analyzer. In both cases, we refer to this data as time distribution curves (TDCs). The $k$-means algorithm sorts the TDCs into $k$ clusters, where $k$ is defined prior to running the code. This results in clusters of TDCs with a similar line shape and it also generates cluster centroids, \emph{i.e.}, the mean TDC for each cluster. This type of analysis has several advantages: it can identify different types of dynamics throughout the entire dataset, it does not depend on prior assumptions or a particular fitting model, and it provides the cluster centroids with a significantly improved signal-to-noise ratio compared to the individual TDCs~\cite{meyerLineShapesTime2025, majchrzakMachinelearningApproachUnderstanding2025a}. 

The TDCs were extracted within a regular grid of ROIs, after Gaussian smoothing of the data. For the FeSuMa data, the smoothing was applied in time (\( \mathrm{FWHM} \simeq 68~\mathrm{fs} \)) and in momentum (\( \mathrm{FWHM} \simeq 0.02~\text{\AA}^{-1} \)). 
For the hemispherical analyzer data, Gaussian smoothing was applied in time (\( \mathrm{FWHM} \simeq 25~\mathrm{fs} \)), in momentum (\( \mathrm{FWHM} \simeq 0.01~\text{\AA}^{-1} \)), and in energy (\( \mathrm{FWHM} \simeq 0.018~\mathrm{eV} \)). 
All smoothing parameters were chosen to remain within the resolution limits of the respective experiment. 
To investigate changes in the dynamics of the excited electrons rather than the absolute photoemission intensity, each TDC was normalized by setting its maximum value to~1 (or its minimum value to $-1$ in case of intensity decreases) prior to applying the clustering algorithm. 

The electronic structure of $R$Te$_3$ near $E_\mathrm{F}$ can be well described by a simple tight-binding model introduced in Ref.~\cite{brouetAngleresolvedPhotoemissionStudy2008} and later expanded to include the possibility of a gap created by an $a$-CDW~\cite{mooreFermiSurfaceEvolution2010} or bilayer splitting to account for the presence of two Te layers in the crystal structure~\cite{chikinaChargeDensityWave2023}. Here, we use the latter version of the model, which corresponds to a $16 \times 16$ Hamiltonian matrix. We shall see that this can explain all the observations, and there is no need to include an $a$-CDW gap. The tight-binding parameters used here are provided in the Appendix.

\section*{Results and Discussion}

Tracking of the time-dependent FC evolution using the FeSuMa analyzer is illustrated in Fig.~\ref{fig:fig1}. Panels (c) and (d) show the photoemission intensity at $E_\mathrm{F}$ before the arrival of the pump pulse and near peak excitation, respectively. Due to the experimental geometry, the absolute photoemission intensity is lower on the left side (negative $k_x$ values). Upon excitation, the melting of the CDW phase can be observed. The closing of the CDW gap is evident in the intensity increase along the $\Gamma$–Z direction. In particular, the inner square FC becomes visible.
When taking the difference between the photoemission intensity at peak excitation and the equilibrium intensity, these changes become even more apparent, as shown in Fig.~\ref{fig:fig1}(e).  Surprisingly, the intensity increase is not restricted to the inner square. Instead, it also appears along the outer FC, which should remain largely unaffected by the $c$-CDW transition aside from effects caused by crossing replica bands. This enhancement is unlikely to arise purely from a hot-electron effect, as shown in the Appendix, where we demonstrate that the dynamics of the populated states dominate the signal.
It is worth noting that a second CDW phase along the $a$-axis would naively be expected to have the opposite effect: it would \emph{reduce} the photoemission intensity along the $k_x$ direction, particularly affecting the states near the X point. The effect of the detailed interaction between the two CDWs is discussed at the end of this section.

For a preliminary look at the dynamics of the CDW melting and the possible emergence of a second CDW, we extract the photoemission intensity as a function of time delay
for two ROIs: one along the inner FC (purple square in Fig.~\ref{fig:fig1}(c)–(e)) and one along the outer FC (yellow square). The resulting TDCs are shown in Fig.~\ref{fig:fig1}(f). In both cases, we observe a sudden intensity increase followed by a slower decay.
As noted above, the intensity increase in the purple ROI, extracted from the inner FC, is expected due to the melting of the $c$-CDW. By contrast, the origin of the intensity increase in the outer FC is not immediately clear. The data further suggest that the signal from the inner FC (purple) decays more rapidly than that from the outer FC (yellow), providing the first evidence for a more complex behavior than the simplified picture in which CDW melting affects only the inner FC.

\begin{figure}[t]
    \includegraphics[scale=0.7]{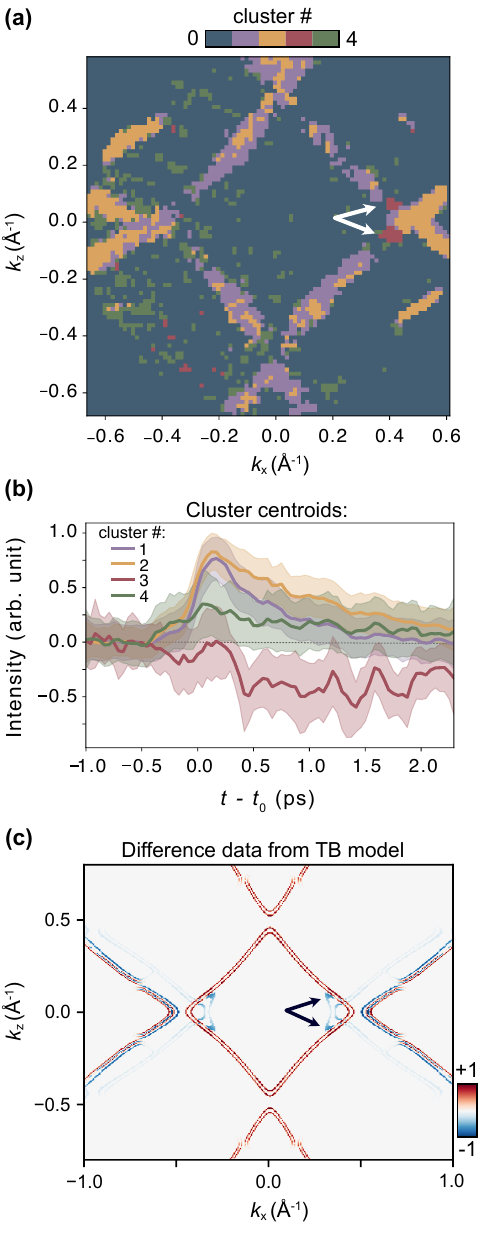}
    \caption{$k$-means clustering ($k = 5$) of the time distribution curves at each point of the Fermi contour, using the data from Fig.~\ref{fig:fig1}(e).
(a) Spatial distribution of the five clusters.
(b) Corresponding cluster centroids.
(c) Tight-binding calculation corresponding to Fig.~\ref{fig:fig1}(e), i.e., the difference between a model without a CDW and one with a fully established $c$-CDW. The color scale is chosen to match that used in the experimental data in Fig. \ref{fig:fig1}(e).
 }
    \label{fig:clusterFesuma}
\end{figure}

Given the complexity of this system, investigating the dynamics across the entire FC becomes essential. The conventional approach of placing multiple ROIs and comparing their TDCs has significant limitations, as it relies on arbitrary choices and risks missing important features. We overcome this by using $k$-means clustering to systematically explore the dynamics across the whole FC.
The clustering analysis, performed with five clusters ($k=5$) for the difference data shown in Fig.~\ref{fig:fig1}(e), is presented in Fig.~\ref{fig:clusterFesuma}(a) and (b). One cluster serves as a background cluster (cluster 0, dark blue), containing all ROIs where the intensity falls below a predefined threshold. Fig.~\ref{fig:clusterFesuma}(a) shows the momentum distribution of the clusters, while Fig.~\ref{fig:clusterFesuma}(b) presents the corresponding cluster centroids.

Cluster 1 (purple) mainly coincides with the inner FC, whereas cluster 2 (yellow) traces the outer FC.  Note that $k$-means clustering is unaware of the spatial locations of the TDCs it processes — such that the spatial separation is therefore an outcome of the clustering itself, indicating that the inner and outer FCs exhibit different dynamics. This is clearly reflected in the cluster centroids in Fig.~\ref{fig:clusterFesuma}(b). Due to the greatly improved signal-to-noise ratio compared to the single-ROI analysis in Fig.~\ref{fig:fig1}, it is now evident that the inner FC indeed recovers to its equilibrium state substantially faster than the outer FC.

Cluster 4 also shows an intensity increase followed by a slow decay upon excitation. However, its distribution in the BZ appears rather random on the left-hand side of the image, where the photoemission intensity is low and the noise level correspondingly high. Moreover, the maximum value reached by the centroid is only $\approx 0.3$. Since all centroids are the mean of TDCs that have been normalized to a maximum (minimum) value of 1 (–1), such a low maximum value combined with the random distribution of corresponding ROIs in $k-$space, indicates substantial variation across the cluster—presumably due to noise.
This issue can, to some extent, be mitigated by increasing the background intensity threshold so that the corresponding TDCs are moved into the background cluster. However, this approach risks also removing meaningful data from the inner FC.

\begin{figure*}[t]
\includegraphics[scale=0.75]{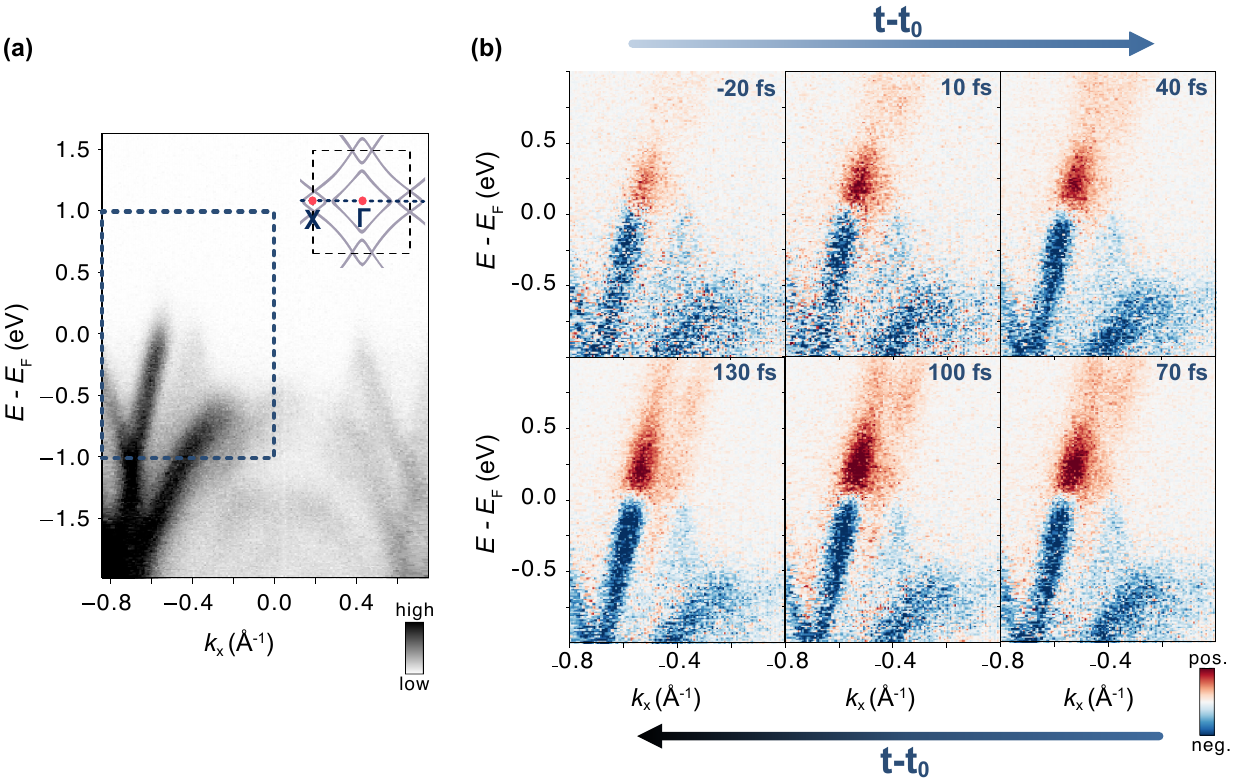}
\caption{Time-resolved photoemission intensity along the $\Gamma$–X direction.
(a) Equilibrium dispersion measured before $t_0$. The inset indicates the orientation of the momentum cut.
(b) Zoomed-in difference plots for a selection of pump–probe delays, corresponding to the area enclosed by the dashed rectangular in (a). }
\label{fig:hemi}
\end{figure*}
An interesting observation is the emergence of a cluster with an intensity loss (cluster 3, red area and TDC) that is mainly confined to two well-defined regions near the butterfly structure, indicated by the white arrows in Fig.~\ref{fig:clusterFesuma}(a). The symmetry-equivalent structures on the left-hand side are not detected:  this is not necessarily because the dynamics there are different from the dynamics on the right-hand side, but because the photoemission intensity on the left-hand side falls below the intensity threshold for the clustering.
The centroid of cluster 3 exhibits strong intensity fluctuations, in contrast to the relatively smooth centroids of clusters 1 and 2. While noise could contribute to this behavior, we will show below that this region of the BZ also displays strong coherent oscillations of the CDW amplitude mode during the re-establishment of the $c$-CDW.

An intensity loss in the region of the BZ covered by cluster 3 is particularly interesting because it could indicate a gap opening, possibly related to the $a$-CDW. However, intensity losses are also expected in the case where only the $c$-CDW is present. This can be seen from the tight-binding model for a single $c$-CDW.
To this end, we present in Fig.~\ref{fig:clusterFesuma}(c) a calculation corresponding to the difference plot of Fig.~\ref{fig:fig1}(e): the broadened FC for LaTe$_3$ without a CDW minus the corresponding calculation for a fully established $c$-CDW. The color scale is chosen such that red and blue correspond to spectral function increases and decreases, respectively. As expected, the most prominent change is the red inner FC, which is removed by the $c$-CDW. Note that the inner FC appears as a double line due to the inclusion of bilayer splitting in the model  \cite{brouetFermiSurfaceReconstruction2004, chikinaChargeDensityWave2023}. This, however, is not resolved in the experiment.  In the region of the butterfly structure we find large, well-defined blue regions corresponding to an intensity loss. 
These regions are well-separated from other bands and do not simply arise from a band shift which leads to parallel red and blue lines, as in parts of the outer FC.
The tight-binding model can therefore qualitatively account for the observation of cluster 3, even though it is known to be insufficient for a quantitative description of the fine structure near the butterfly feature \cite{chikinaChargeDensityWave2023}.
An open question is whether the intensity increase in the outer FC and its slower decay compared to that of the inner square can likewise be explained within a model based solely on a single $c$-CDW. 

To investigate the situation along the $k_x$ direction in more detail, we complement the FeSuMa results with data collected using a hemispherical analyzer. This configuration extends the accessible energy range beyond the states immediately around $E_\mathrm{F}$.
Figure~\ref{fig:hemi} presents such a dataset. Panel (a) shows the dispersion measured prior to the arrival of the pump laser at $t_0$. The state carrying a large spectral weight and crossing $E_\mathrm{F}$ at approximately $k_x = -0.6$~\AA$^{-1}$ is the tip of the outer FC (blue in Fig.~\ref{fig:fig1}(a)), while the feature with its maximum just below $E_\mathrm{F}$ at about $k_x = -0.4$~\AA$^{-1}$ originates from the bands forming the butterfly structure.

Figure~\ref{fig:hemi} (b) shows the zoomed-in difference data within the dashed area marked in panel (a), for several pump–probe delays. After excitation, the unoccupied states just above $E_\mathrm{F}$ are populated by hot electrons up to $E - E_\mathrm{F} \approx 1$~eV. Up until about 40~fs after the excitation, the intensity of the previously unoccupied bands exhibits a local minimum at $E - E_\mathrm{F} \approx 0.5$~eV. At later delay times, however, this intensity drop disappears, and by $t - t_0 = 130$~fs two continuous bands are crossing $E_\mathrm{F}$.

\begin{figure}[!t]
    \includegraphics[scale=1]{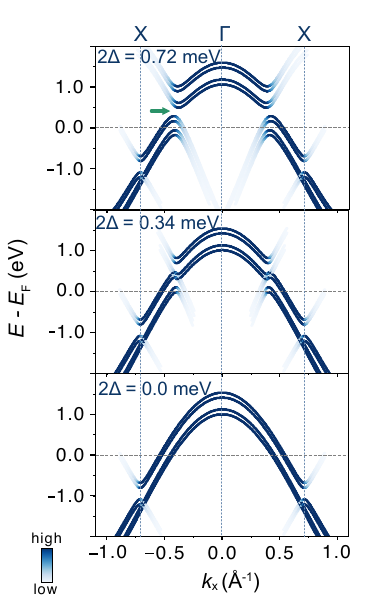}
    \caption{Tight-binding model calculations for varying gap sizes $2\Delta$ along the $k_x$ direction. The gap is indicated by the green arrow.
(top) Electronic structure in the fully developed $c$-CDW phase.
(middle) Intermediate stage with a reduced gap.
(bottom) Normal phase with no CDW gap.
Only the spectral weight of the original $p_x$ and $p_z$ orbitals is shown. }
    \label{fig:TB}
\end{figure}

These findings can be explained by the tight-binding model calculations shown in Fig.~\ref{fig:TB}. The top panel depicts the electronic structure in the CDW phase with a gap size of $2\Delta = 0.72$~meV (indicated by the green arrow), which is close to the experimentally determined value in equilibrium \cite{huCoexistenceCompetitionMultiple2014}. The middle panel shows the same calculation but with a reduced gap size of $2\Delta = 0.34$~meV, and the bottom panel shows the normal state with no CDW gap ($2\Delta = 0$~meV). Again, the calculated bands are split due to the bilayer interaction.
The calculations closely reproduce the experimental results in Fig.~\ref{fig:hemi}(b). They indicate that the $c$-CDW gap, which removes the FC along the $k_z$ direction, is also present along $k_x$, but appears at much higher energies above $E_\mathrm{F}$. This gap progressively closes following light excitation, resulting in two (doubled) bands crossing $E_\mathrm{F}$ in the normal state.

In the experimental data, the situation is more complex than in Fig.~\ref{fig:TB} because intensity changes do not only arise from band shifts in energy and $k$ but also from the generation of hot electrons and holes. Moreover, the slow re-opening of the $c$-CDW gap is not a continuous process but is accompanied by strong coherent oscillations of the CDW's amplitude mode, as we shall see below. Nevertheless, the qualitative agreement between the calculations and the data is excellent. 
The calculations in Fig.~\ref{fig:TB} also provide a straightforward explanation for the slower dynamics observed in cluster 2 compared to cluster 1 in Fig.~\ref{fig:clusterFesuma}(b): all of the intensity changes discussed so far are ultimately tied to the gap parameter of the $c$-CDW. The key difference between the inner and outer FCs is that the center of the $c$-CDW gap lies along  the $\Gamma-Z$ direction and the gap is symmetric around $E_\mathrm{F}$. During the re-formation of the CDW, even a small gap parameter is sufficient to remove states near $E_\mathrm{F}$ for the inner FC.
The situation for the outer FC is entirely different. When $2\Delta$ increases from zero to finite values, a gap first develops at high energies (see green arrow in Fig.~\ref{fig:TB}) in the unoccupied states, rather than at $E_\mathrm{F}$. Intensity changes near $E_\mathrm{F}$ only begin to appear later, when $2\Delta$ becomes larger. Thus, different dynamics can naturally be expected across the FC—even in a system governed by a single order parameter.

\begin{figure}[t]
    \includegraphics[scale=1]{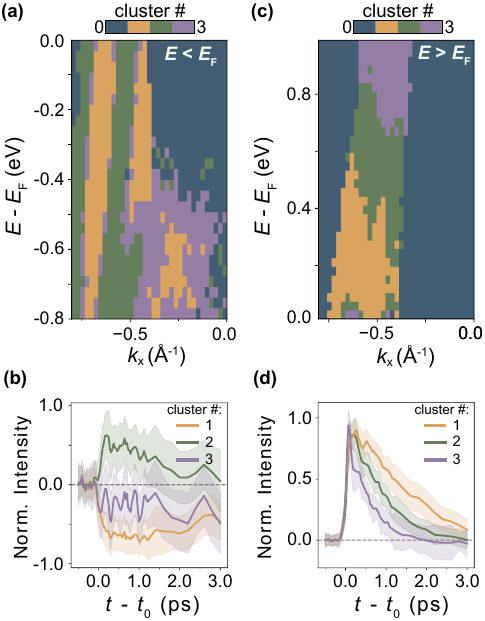}
    \caption{Clustering analysis of the normalized time distribution curves extracted from the data in Fig.~\ref{fig:hemi} using $k = 4$ clusters.
(a) Cluster distribution for the states below $E_\mathrm{F}$.
(b) Corresponding cluster centroids, \emph{i.e.}, the mean time distribution curve of each cluster after subtraction of a constant background.
(c), (d) Same analysis for the states above $E_\mathrm{F}$.
 }
    \label{fig:clusterhemi}
\end{figure}

\begin{figure*}[t]
\includegraphics[scale=1]{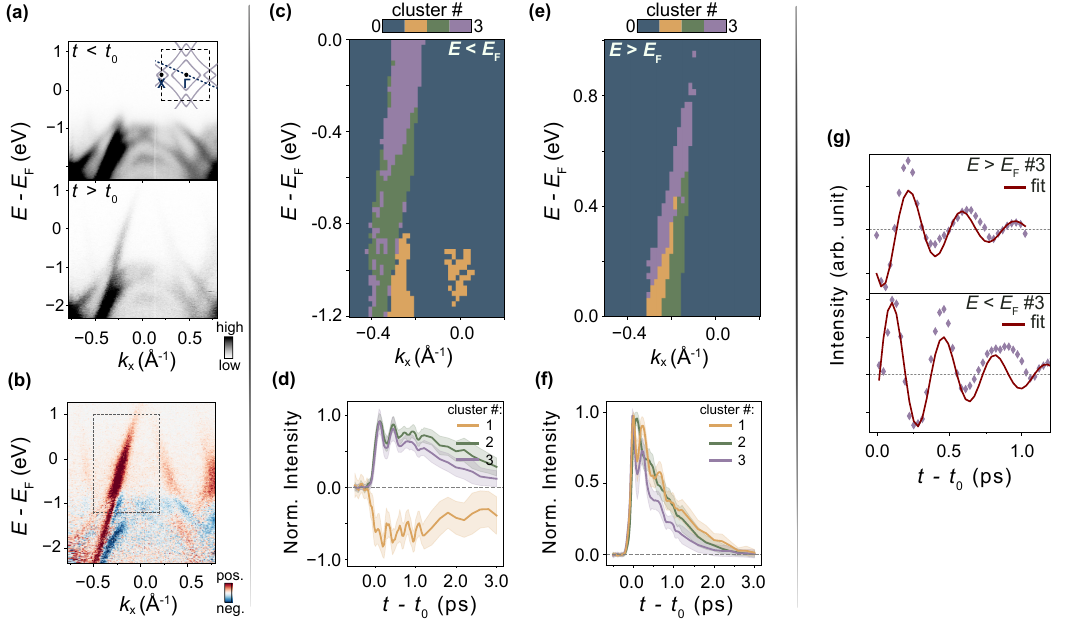}
\caption{Clustering of TR-ARPES data along a direction containing only the inner square Fermi contour crossing.
(a) Top: Data in equilibrium.
 Bottom: Data close to peak excitation ($t - t_0 = 130$fs).
(b) Difference data for the data shown in (a). The black dashed lines mark the relevant regions for the clustering.
(c) Cluster distribution for the states below $E_\mathrm{F}$.
(d) Corresponding cluster centroids.
(e) Cluster distribution for the states above $E_\mathrm{F}$.
(f) Corresponding cluster centroids.
(g) Oscillations in the time distribution curves after subtraction of an exponentially decaying background for cluster 3 of the unoccupied states (top) and cluster~3 of the occupied states (bottom). A fit to the data is shown by the red solid line.
}
\label{fig:clusterhemi2}
\end{figure*}

We further examine the situation along the $k_x$ direction by applying $k$-means clustering to the dataset collected with the hemispherical analyzer. The results are shown in Fig.~\ref{fig:clusterhemi}. We cluster the TDCs extracted from the dispersion in Fig.~\ref{fig:hemi} using $k = 4$ clusters, including a background cluster that contains all TDCs with intensities below a predefined threshold (dark blue).
The clustering is performed separately for occupied and unoccupied states, excluding the region very close to $E_\mathrm{F}$, where the intensity of the difference data are essentially zero and thus dominated by noise (see the sharp transition between predominantly blue and predominantly red regions in Fig.~\ref{fig:hemi}(b)). This approach also allows the background intensity to be adjusted independently for the two clustering regions, since the measured signal in the unoccupied states is much weaker and therefore requires a lower intensity threshold. The cluster maps and centroids for the occupied (unoccupied) states are presented in Figs.~\ref{fig:clusterhemi}(a) and \ref{fig:clusterhemi}(b) (Figs.~\ref{fig:clusterhemi}(c) and \ref{fig:clusterhemi}(d)).

For the occupied states, clusters 1 and 2 display strikingly opposite behavior. This is readily explained by the two bands near $E_\mathrm{F}$ shifting slightly in $k_x$, as also predicted by the tight-binding model. The dynamics is completely explained by such a shift: cluster 2 gains intensity at the rate cluster 1 looses intensity. The absence of any additional features in the dynamics supports the conclusion that inclusion of an $a$-CDW is not necessary to explain our data.

For the unoccupied states, we observe the expected energy-dependent dynamics: the cluster located closest to $E_\mathrm{F}$ (yellow) exhibits the slowest decay \cite{meyerLineShapesTime2025}. Furthermore, the high signal-to-noise ratio in the cluster centroids clearly reveals  signatures of oscillations in the measured intensity as a function of delay time for both the occupied and unoccupied states.

Finally, the centroids for the purple clusters in Figs.~\ref{fig:clusterhemi}(a) and \ref{fig:clusterhemi}(b) reach much lower magnitude in the intensity than the other centroids, remaining far below the normalization values of 1 and –1 of a single TDC. This again suggests that these clusters contain a greater variety of dynamics than the others.

As a comparison, we apply the $k$-means analysis to a $k$-space direction with a less complex electronic structure and dynamics. Figure~\ref{fig:clusterhemi2} presents the corresponding analysis for a direction that contains only the inner square FC. The raw photoemission intensity is shown in Fig.~\ref{fig:clusterhemi2}(a)  for data at equilibrium and close to peak excitation ($t - t_0 = 130$~fs). The corresponding difference plot is given in Fig.~\ref{fig:clusterhemi2}(b). The results of the clustering analysis are displayed in Figs.~\ref{fig:clusterhemi2}(c)–\ref{fig:clusterhemi2}(f). The clusters around $E_\mathrm{F}$ show strong intensity gains upon excitation, followed by a slow decay—behavior consistent with the closing of the gap around the inner FC followed by a slow re-opening. The situation is still not completely trivial, as it reflects both the band structure changes and the creation and decay of hot carriers, but it is nevertheless simpler than in Fig.~\ref{fig:clusterhemi} because only a single band is involved.

The coherent oscillations observed in the cluster centroids of Fig.~\ref{fig:clusterhemi2} are much more pronounced than in Fig.~\ref{fig:clusterhemi}. To investigate these oscillations in more detail, we focus on the centroids of cluster 3 in the occupied- and unoccupied-state data (purple areas and TDCs in Fig.~\ref{fig:clusterhemi2}(c-f)), as these clusters exhibit strong oscillations.

The oscillations in the intensity after subtraction of an exponentially decaying background are shown in Fig.~\ref{fig:clusterhemi2}(g). The oscillations for the unoccupied states are plotted in the top panel. Fitting these oscillations to a damped cosine function within the range of 0-1~ps (red solid line) yields a period of $T = 354 \pm 11$~fs, corresponding to a frequency of $f = 2.8 \pm 0.1$~THz, in good agreement with literature values for the amplitude mode of the $c$-CDW \cite{maklarCoherentModulationQuasiparticle2022, liuDistinctAmplitudeMode2024, Schmitt:2008aa, rettigPersistentOrderDue2016}.
For the intensity variations extracted from the occupied states (bottom panel), a single periodic fit cannot adequately reproduce the oscillatory behavior. This is evident in the full range fit shown in the bottom panel of  Fig.~\ref{fig:clusterhemi2}(g). While this fit likewise yields a frequency of $f = 2.8 \pm 0.1$~THz, the fit begins to deviate substantially from the data at times beyond approximately 0.9~ps, where the period of the oscillation appears to increase.
Such an apparent phonon mode-softening at later time delays ($>1$~ps)  agrees with recent observations for pump energies that exceed the CDW gap size, as is the case here \cite{liuDistinctAmplitudeMode2024}. 

We stress that the detailed mechanism leading to the oscillations in the TDCs is complex, as it arises from a combination of the CDW amplitude mode and a simultaneous band shift. This can also be seen in the data: A band gap opening upon the recovery of the CDW would imply that the band ``bends downwards'' and thus shifts intensity from the purple cluster to the green cluster in Fig.~\ref{fig:clusterhemi2}(c) and (d). Such an intensity shift is indeed observed in the cluster centroids. Moreover, the ``bending down'' scenario is supported by the steep boundaries between the cluster locations in Fig.~\ref{fig:clusterhemi2}(c). A simple dynamics dictated by the Fermi-Dirac distribution would lead to horizontal boundaries between clusters but a ``bending down'' shifts intensity side-ways as well (even though resolution effects can modify this simple scenario \cite{meyerLineShapesTime2025}). We therefore stress that a comprehensive description of the detailed dynamics requires a more detailed analysis.

\begin{figure}[t]
\includegraphics[scale=1.5]{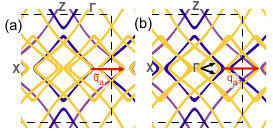}
\caption{Nesting conditions for different nesting vectors along $k_x$.
(a) The nesting vector $\tilde{q}_a$ has roughly the same length as the $c$-CDW nesting vector $q_c$.
(b) The nesting vector $q_a$ is about 10\% longer than $\tilde{q}_a$ shown in panel (a). The black arrows in (b) indicate the position of the butterfly structure in the $c$-CDW.}
\label{fig:nestingfigure}
\end{figure}

Finally, we return to the question of why we do not observe any clear signatures of a transient $a$-CDW—neither directly in the time-dependent FC and band structure, nor in coherent oscillations that could potentially indicate a second amplitude mode. Guided by the fact that the $a$-CDW in the heavier $R$Te$_3$ compounds has a relatively small gap and only a minor effect on the Fermi contour \cite{mooreFermiSurfaceEvolution2010}, one possible conclusion is that our experiment lacks the required sensitivity. However, this small-gap scenario appears rather unlikely.

To see this, we consider Fig.~\ref{fig:nestingfigure}, which schematically illustrates the Fermi contour for two different nesting scenarios along $k_x$ (with the $c$-CDW ignored). In panel (a), the yellow FC replicas generated by translating the FC with the vector $\tilde{q}_a$ nest almost perfectly with the corners of the inner square in the $k_x$ direction. This is essentially the same situation as for the $c$-CDW, but rotated by $90^\circ$. It would thus result in a large portion of the FC being gapped out. In panel (b), by contrast, $q_a$ is about 10\% larger than $q_c$ — this corresponds to the case actually found for the equilibrium $a$-CDW of other $R$Te$_3$ compounds \cite{Maschek:2018aa}. In this situation, the nesting is far from perfect resulting in the smaller gaps observed for these compounds \cite{mooreFermiSurfaceEvolution2010}.

Combining the sketch in Fig.~\ref{fig:nestingfigure}(b) with the tight-binding model in Fig.~\ref{fig:clusterFesuma}(c), we can understand why the longer $q_a$ is preferred over the shorter $\tilde{q}_a$ when the $c$-CDW is already present. In this case, the $c$-CDW already gaps most of the inner square FC. This is seen by the red inner square in Fig.~\ref{fig:clusterFesuma}(c), which corresponds to spectral weight that is present in the undistorted lattice but removed by the $c$-CDW. In the vicinity of the X points, the $c$-CDW leads to a more complex re-arrangement of the bands, giving rise to the butterfly structure marked by the arrows in Fig.~\ref{fig:clusterFesuma}(c). The butterfly structure appears in blue in the difference plot, meaning that is is \emph{created} by the $c$-CDW. The emergence of the butterfly and simultaneous disappearance of the inner square near X corresponds to a shift of spectral weight to a lower $k_x$. The $a$-CDW can now potentially gain energy by gapping the region around the butterfly. Achieving this requires the longer $q_a$ nesting vector, as this creates the yellow replica bands in Fig.~\ref{fig:nestingfigure}(b) at a lower $k_x$ than in Fig.~\ref{fig:nestingfigure}(a). Indeed, the position of the replica bands near the X point in Fig.~\ref{fig:nestingfigure}(b), see black arrows, matches fairly well with the location of the butterfly structure in Fig.~\ref{fig:clusterFesuma}(c). In the absence of a $c$-CDW, on the other hand, one would expect nesting with a vector $\tilde{q}_a$, as in Fig.~\ref{fig:nestingfigure}(a), as this would give rise to the largest gap. This is consistent with experimental observation of a short $q_a \approx q_c$ for the transient CDW state \cite{Kogar:2019ab}. It is also in line with the picture of the transient $a$-CDW nucleation being facilitated by topological defects, assuming that the $a$-CDW is growing in regions free of $c$-CDW \cite{Kogar:2019ab} . Since these considerations imply a large gap for the transient $a$-CDW, its contributions should  be detectable in our measurements and in particular in Fig.~\ref{fig:hemi}.

Another possible reason for the absence of any signatures of the transient $a$-CDW here could be that this state never achieves long-range order before it is removed by the re-establishment of the static $c$-CDW. Zong \emph{et al.} estimate the correlation length of the transient $a$-CDW to be between 3.5 and 10~nm~\cite{Zong:2021aa}. This is orders of magnitude below the coherence length of the equilibrium $c$-CDW but not short enough to preclude the observation of its effect in electron diffraction.

\section*{Conclusion}

We have investigated the out-of-equilibrium electronic structure of LaTe$_3$ using TR-ARPES, combining conventional analysis techniques with $k$-means clustering. The $k$-means approach has proven to be a powerful tool for analyzing large datasets, allowing the dynamics to be examined at every point along the Fermi contour. The resulting cluster centroids exhibit substantially improved signal-to-noise ratios compared to single-ROI TDCs, revealing not only distinct decay times for the inner and outer Fermi contours but also small regions that display electron-depletion behavior.

Our analysis suggests that the slower decay time observed for electrons in the outer Fermi contour originates from the specific location of the CDW gap far above the Fermi level in the $k_x$ direction, in contrast to the $k_z$ direction, where the gap is symmetric around $E_\mathrm{F}$. Because of these different gap locations, the gap on the inner FC re-opens immediately upon re-formation of the CDW, whereas the photoemission intensity of the outer FC is only affected once the order parameter reaches sufficiently large values.
The application of clustering analysis to hemispherical-analyzer data also yields cluster centroids with a high signal-to-noise ratio, revealing clear oscillations in the measured intensity that can be attributed to the coherent amplitude mode of the CDW. This, in turn, enabled a detailed analysis of the oscillations.

Despite recent ultrafast electron diffraction experiments suggesting the presence of a light-induced second CDW phase perpendicular to the equilibrium CDW, our detailed electronic-structure measurements reveal no detectable signatures of such a phase. This conclusion is supported by both FC measurements using the FeSuMa analyzer and electronic-dispersion measurements obtained with conventional hemispherical analyzers. Our results therefore indicate that, while structural changes may occur upon optical excitation, they do not manifest as a distinct secondary CDW phase in the electronic structure of LaTe$_3$. This highlights the importance of employing complementary spectroscopic techniques to fully understand light-induced phase transitions in quantum materials.

The data of this study is available at \cite{dataref}.

\section*{Acknowledgements}
We thank Uwe Bovensiepen and Florian Denizer for helpful discussions. 
This project has received funding from the European Union's Horizon 2020 research and innovation programme under grant  agreement no. 871124 Laserlab-Europe. We gratefully acknowledge funding from Independent Research Fund Denmark  (Grant No. 1026-00089B and Grant No. 4258-00002B) and Novo Nordisk Foundation (Grant number NFF23OC0085585).
Work at Princeton was supported by the Air Force office of Scientific Research under award number FA9550-24-1–011.

\vspace{1cm}
\section{Appendix}

\subsection{FeSuMa clustering}

In Fig.~\ref{fig:clusterFesuma} we present the $k$-means clustering result of the FC measured using a FeSuMa analyzer. In Fig.~\ref{fig:clusterFesuma} we present the results following the same clustering procedure, but for a retarding voltage set 100~meV below $E_{\text{F}}$. Otherwise the sample and measurement conditions are identical. Comparing this data set to the one discussed in Fig.~\ref{fig:clusterFesuma} -- collected with a retarding voltage 25~meV below $E_{\text{F}}$ -- should enable us to judge the degree to which the presence of hot electrons influences our results. 

The reasoning behind this is as follows: As explained in the main text, the FeSuMa's retarding field permits all electrons with a kinetic energy above a threshold to be detected. In equilibrium, it thus maps the Fermi contour when the retarding voltage is set to a value just below $E_{\text{F}}$. In the presence of hot electrons after exciting the sample, this is not necessarily the case: For a sufficiently high density of hot electrons, the measured contours could become washed out or shifted since the hot electrons follow the unoccupied dispersion and are thus present at different momenta. In the main paper we argued that the effect should be small because of the relatively small number of hot electrons. Here we test this: re-taking the data a retarding voltage 100~meV below $E_{\text{F}}$ greatly enhances the signal from the occupied states and thus suppresses the relative contribution of the hot electrons. This comes at the expense of potential broadening and shifting because now more occupied states contribute to the signal, and these are also at different momenta than the electrons at the FC. However, comparing the clustering results here and in the main paper shows that they are essentially the same, from which we conclude the influence of hot electrons is negligible. Indeed, extending the detected energy range by 75~meV into the occupied states does not cause a major change of the resulting time distribution curves, either. One might expect because the CDW gap is still much wider than the total width of the energy window detected by the FeSuMa.

\begin{figure}[t]
    \includegraphics[scale=0.7]{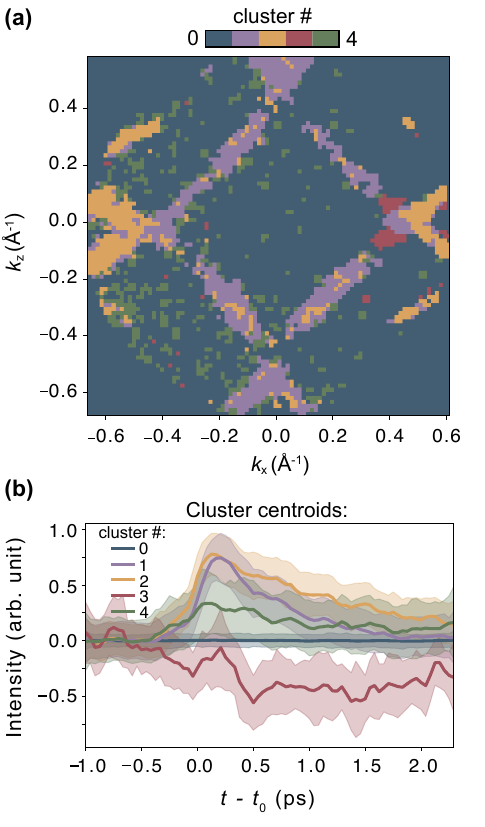}
    \caption{$k$-means clustering ($k = 5$) of the time distribution curves from a data set corresponding to that of Figs.~\ref{fig:fig1} and \ref{fig:clusterFesuma}  but with a retarding voltage set to 100~meV below $E_\mathrm{F}$.
(a) Spatial distribution of the five clusters.
(b) Corresponding cluster centroids.
 }
    \label{fig:clusterFesuma}
\end{figure}

\subsection{Tight-binding calculations}

The tight binding model used in this study is based on the calculations introduced in Ref.~\cite{brouetAngleresolvedPhotoemissionStudy2008} and further refined in Ref.~\cite{chikinaChargeDensityWave2023}. Here we use the same model as in the latter work but we adapt the parameters for LaTe$_3$. These parameters are given in Table~\ref{t1}. 

\begin{table}
\centering
\begin{tabular}{lc}
\hline
Parameter & Value \\
\hline
$a$            & $4.42$ \AA \\
$c$            & $4.42$ \AA \\
$q_c$	&        $0.68\,c^*$\\
$t_{\parallel}$ & $-1.5$ eV \\
$t_{\perp}$    & $0.3$ eV \\
$t_{\mathrm{int}}$ & $0.2$ eV \\
$V$            & $0.36$ eV \\
$f$            & $0.10$ eV \\
$t_{\mathrm{bl}}$ & $0.03$ eV \\
$V'$           & $0.16$ eV \\
\hline
\end{tabular}
\caption{Tight-binding parameters used for the calculations shown in the main manuscript.}
\label{t1}
\end{table}

%

%

\end{document}